\documentclass{ws-mpla}
\usepackage[super]{cite}
\usepackage{graphicx}
\usepackage{dcolumn}
\usepackage{bm,epstopdf}
\usepackage{epsfig,amsmath}
\usepackage[usenames,dvipsnames]{color}
\usepackage[linktocpage,pagebackref]{hyperref}
\usepackage{appendix}
\usepackage{textcomp}
\usepackage{float}
\usepackage{subfigure}
\begin{document}


\catchline{}{}{}{}{}{}

\title{Geodesic stability and Quasi normal modes via Lyapunov exponent for Hayward Black Hole }

\author{Monimala Mondal}
\address{Department of Mathematics, Jadavpur University, Kolkata 700032, West Bengal, India
 E-mail-  monimala.mondal88@gmail.com}

\author{Parthapratim Pradhan}
\address{Department of Physics, Hiralal Mazumdar Memorial College For Women,
Dakshineswar, Kolkata-700035, India\\
 E-mail-  pppradhan77@gmail.com }

\author{Farook Rahaman}
\address{Department of Mathematics, Jadavpur University, Kolkata 700032, West Bengal, India
 E-mail- rahaman@associates.iucaa.in}

 \author{Indrani Karar}
 \address{Department of Registrar, Kalyani  University, Nadia, West Bengal, India
 E-mail- indrani.karar08@gmail.com}

\maketitle

\pub{Received (Day Month Year)}{Revised (Day Month Year)}

\begin{abstract}
We derive { proper-time }Lyapunov exponent $(\lambda_{p})$ and
 {coordinate-time } Lyapunov exponent $(\lambda_{c})$ for a
regular Hayward class of black hole.  {The proper-time corresponds to  $\tau$
and the coordinate time corresponds to $t$. Where $t$ is measured by the asymptotic
observers both for   for Hayward black hole and for special case
of Schwarzschild black hole}.
 { We compute }
their ratio as $\frac{\lambda_{p}}{\lambda_{c}} = \frac{(r_{\sigma}^{3} +
2 l^{2} m )}{\sqrt{(r_{\sigma}^{2} + 2 l^{2} m )^{3}- 3 m r_{\sigma}^{5}}}$
 {for} time-like geodesics.  {In the limit of
$l=0$ that means }for Schwarzschild black hole this ratio reduces to
$\frac{\lambda_{p}}{\lambda_{c}} = \sqrt{\frac{r_{\sigma}}{(r_{\sigma}-3 m)}}$.
Using Lyponuov exponent, we investigate the stability and instability of
equatorial circular geodesics. By  {evaluating} the Lyapunov exponent, which is
the inverse of the instability time-scale, we show that, in  {the} eikonal limit,
the real and imaginary parts of quasi-normal modes~(QNMs)
is specified by
the frequency and instability time scale of the null circular geodesics.
 {Furthermore, we discuss the unstable photon sphere and radius of shadow
for this class of black hole.}

\end{abstract}


\section{Introduction}

An elementary set of unstable circular orbits about a Schwarzschild black hole~(BH) are
consequences of the  {non-linearity} of general theory of relativity.
Their instability   {could} be measured by a positive
Lyapunov exponent~\cite{Neil62}. Albeit the Lyapunov
exponent are often related with chaotic dynamics~\cite{Motter03,Dorfman99}, the
geodesics about a Schwarzschild BH are not chaotic; the orbits are
completely solvable and hence integrable. The unstable geodesic orbits should
have positive Lyapunov exponent~\cite{Lyapunov92}, which has the invariant
properties first established in~\cite{Karas92}. Lyapunov exponent has a great
impact on general relativity for its numerous applications: they are relative
and depend on the coordinate system used, they vary from orbit to orbit. In
this work,  {we are interested to focus} on analytical formulation
of Lyapunov exponent and QNMs in terms of the expressions of the radial equation of
circular geodesics about a BH space-time. In this regard an equatorial
circular geodesics about a BH may play  crucial role in general theory
of relativity for classification of the orbits.

 {\noindent~Black holes and singularities are approached to be unavoidable predictions
of the theory of general relativity (GR). In order to solve the black hole singularity,
 several phenomenological propositions have been studied in the existing literatures. Bardeen BH~\cite{Bardeen68}
 was the first model that has proposed as a spherically symmetric compact object with an event
 horizon and satisfying the weak energy condition. In the year 2006, Hayward  \cite{Hayward06} proposed the formation
 and evaporation of a new kind of regular solution in space-time. The static region of a
 Hayward space-time is similar to Bardeen black hole. In the article~\cite{Lin13} , the authors discussed the massive scalar quasinormal modes of the Hayward
  black hole. In this study, variations of the Hayward solutions have also been studied as Hayward with charge~\cite{Frolov16} and Rotating Hayward~\cite{Muhammed15}}.

\noindent~The authors in~\cite{Cardoso09,Cardoso67,Sperhake08} derived the Lyapunov
exponent to study the instability for the circular null geodesics in terms of the
expressions for QNM of spherically symmetric space-time. Here, the main focus
is on the null circular geodesics, which play an important role for the Lyapunov
exponent. The null circular geodesics is described by the shortest possible orbital
period as estimated by the asymptotic observers~\cite{Hod11}. The  time circular geodesics
provide the slowest way to circle the BH, among all the possible circular geodesics.

\noindent The QNMs spectrum for stable BH represents an infinite set of
complex frequencies, which designates damped oscillations for the amplitude. It is
clear that the oscillations with
lower level of damping rate is controlled with slow time, whereas oscillations with higher
level of damping rate are exponentially terminated~\cite{Konoplya11}.
In 1971,  {Press}~\cite{Press71},
first introduced the term `quasi-normal frequency'.  {He showed that
when a Schwarzschild BH is perturbed it vibrates with an angular frequency
$\omega=\frac{\ell}{\sqrt{27}M}$. Where $M$ is BH mass and $\ell$  is
spherical symmetric index. He then interpreted it  as vibration frequency
of BHs.}  In the same year the lowest QNMs
were computed by investigating test particle falling about Schwarzschild BH.
 {So far} the QNMs has been explored substantially in diverse field.
The WKB method gives an exact estimation of QNM frequency in the eikonal limit.
Also WKB method was first introduced by Schutz and Will~\cite{Schutz33} to
analyze the problem of scattering about BH.

\noindent The plan of the ar {proper time} Lyapunov
~\footnote{ {Since the Lyapunov exponent is explicitly coordinate
dependent and therefore have a degree of un-physicality. That's why
we define two types of Lyapunov exponent. One is coordinate type Lyapunov exponent
and the other one is proper time Lyapunov exponent. This is strictly for timelike
geodesics.}}
exponent~$\lambda_{p}$ and
coordinate time Lyapunov exponent ~$\lambda_{c}$ in terms of second order
derivative of the effective potential for the radial motion $\dot{r}^{2}$:
 {\begin{eqnarray}
\lambda_{p} &=& \pm  \sqrt{\frac{(\dot{r}^{2})''}{2}}, \\
\lambda_{c} &=& \pm  \sqrt{\frac{(\dot{r}^{2})''}{2\dot{t}^{2}}}.
\end{eqnarray}}

In  Sec.~3, we describe the equatorial circular geodesics  { of
spherically  symmetric regular Hayward BH. Then we calculate the Lyapunov exponent in terms
of the  timelike circular geodesics and null circular geodesics.  We also discuss the
gravitational bending of light and photon sphere for this BH. }

In Sec.~4,  we derive the relation  between QNMs~ { in the
eikonal limit and  Lyapunov exponent of a static,  spherically symmetric  regular
Hayward BH.  In the limit $l=0$, one obtains the QNMs frequency of spherically-symmetric
Schwarzschild BH which was first calculated in~\cite{Pradhan56}}.

~ {Moreover we compute the angular velocity $\Omega_{c}$ of the unstable null geodesics.}
~ {Also, we compute  the Lyapunov exponent, which investigates
the instability of the time-scale of circular orbit}~\cite{Cornish03,Bombelli92,Moni19}
~ {also which is } in agreement with analytic WKB approximations ~ {of}
QNMs~ {of the Hayward  BH in the eikonal limit. Thus the QNMs frequency in the eikonal
limit is found to be}
\begin{eqnarray}
\omega_{QNM} &=& j\, \bigg( \frac{\sqrt{m\, (r_{c}^{3} - 4m\, l^{2} )}}{(r_{c}^{3} + 2m\,l^{2})}\bigg)
- i \left( n +\frac{1}{2}\right) \sqrt{\frac{3m^2 [ r_{c}^{5} -6 l^{2} ( r_{c}^{3} -2m\,l^{2}  )]}
{ (r_{c}^{3} + 2m\, l^{2} )^3}},
\end{eqnarray}
where $ n$ represents the overtone number and $j$ represents the angular momentum of
the perturbation.

For schwarzschild BH, the QNMs frequency becomes
\begin{eqnarray}
\omega_{QNM} &=& j \sqrt{\frac{m}{r_{c}^{3}}} - i \left(n + \frac{1}{2} \right)\frac{\sqrt{3}m}{r_{c}^{2}}.
\end{eqnarray}
The real part of the complex QNM frequency can be determined by the angular velocity of the
unstable null geodesics and the imaginary part is related to the instability time scale of
the orbit. Finally, we briefly discuss about the outcome of this paper
~ {in Sec.~5}.

\section{Proper time Lyapunov exponent, ~{Coordinate time Lyapunov exponent}
and Geodesic stability}
The Lyapunov exponent or Lyapunov characteristic exponent of a dynamical
system is a measure of the average rate of expansion and contraction of adjacent trajectories in the
phase space. A negative Lyapunov exponent designates the convergence between nearby trajectories.
A positive Lyapunov exponent determines the divergence between nearby geodesics in which the path
of such a system are the most active to change the starting circumstances. The vanishing
Lyapunov exponent designates the existence of marginal stability. ~ { The equation of motion
in terms of Lyapunov exponents for geodesic stability analysis  should  be} written as
\begin{eqnarray}
\frac{dZ_{i}}{dt}=F_{i}(Z_{j}),
\end{eqnarray}
and its linearized form around a certain orbit is
\begin{eqnarray}
\frac{d \delta Z_{i}(t)}{dt}= A_{ij}(t) \delta Z_{j}(t).
\end{eqnarray}
Here,
\begin{eqnarray}
A_{ij}(t)= \frac{\partial F_{i}}{\partial Z_{j}}\Bigg|_{Z_{i}(t)},
\end{eqnarray}
represents the linear stability matrix~\cite{Cornish03}. Now, the solution of
the Eq.~(2) can be written as
\begin{eqnarray}
\delta Z_{i}(t)= X_{ij}(t) \delta Z_{j}(0),
\end{eqnarray}
where $X_{ij}(t)$ represents the evolution matrix, which leads to
\begin{eqnarray}
\dot{X}_{ij}(t) &=& A_{im} X_{mj}(t),
\end{eqnarray}
with $ X_{ij}(0) = \delta_{ij}$. The principal Lyapunov exponent $\lambda$ can be
expressed in terms of the eigenvalues $X_{ij}$ as follows:
\begin{eqnarray}
\lambda &=& \lim_{t\rightarrow \infty} \frac{1}{t} \log \bigg(\frac{ X_{ij}(t)}{ X_{mj}(0)} \bigg).
\end{eqnarray}
If there exists a set of $n$ Lyapunov exponents connected with an n-dimensional
independent system, then they can be arranged by the size as
\begin{eqnarray}
\lambda_{1}\geq\lambda_{2}\geq\lambda_{3}\geq,.....,\geq\lambda_{n}.
\end{eqnarray}
The set of $n$~numbers of $\lambda_{i}$ are known as Lyapunov spectrum.\\

\noindent In an equatorial plane, for any static spherically symmetric
space-time, the Lagrangian  {of a }  test particle
can be written as
\begin{eqnarray}
\mathcal{L} = \frac{1}{2}\bigg[ g_{tt}\,\dot{t}^{2} + g_{rr}\,\dot{r}^{2} + g_{\phi\phi}\,\dot{\phi}^{2} \bigg].
\end{eqnarray}
From the above expression, the canonical momenta can be derived as
\begin{eqnarray}
\label{Mom}
p_{q} = \frac{\partial \mathcal{L}}{\partial\dot{q}}.
\end{eqnarray}
The generalized momenta derived from the above Lagrangian are
 \begin{eqnarray}
p_{t}& =& g_{tt} \dot{t} = - E = \textrm{Const},\\
p_{\phi}& =& g_{\phi\phi} \dot{\phi} = L = \textrm{Const},\\
p_{r} &= &g_{rr} \dot{r}.
\end{eqnarray}
Here, `dot' represents the differentiation with respect to proper time $(\tau)$.
E and L are the energy and angular momentum per unit rest mass of the test
particle,  respectively.  \\

\noindent Now, from Euler-Lagrange's equation of motion, we can write
\begin{eqnarray}
\frac{d p_{q}}{d\tau} = \frac{\delta \mathcal{L}}{\delta q}.
\end{eqnarray}
Linearizing the above equation of motion in two-dimensional phase space with
respect to $Z_{i}(t) = (p_{r},r)$, around the circular
orbit (taking $r$ as a constant), we get
\begin{eqnarray}
\frac{d p_{r}}{d\tau} = \frac{\partial \mathcal{L}}{\partial q}~~~~
\textrm{and}~~~~  \frac{dr}{d\tau} = \frac{p_{r}}{g_{rr}},
\end{eqnarray}
and an infinitesimal evolutionary matrix can be expressed as
\begin{eqnarray}
A_{ij} = \left(
           \begin{array}{cc}
             0 & A_1 \\
             A_2 & 0 \\
           \end{array}
         \right),
\end{eqnarray}
where
\begin{eqnarray}
A_{1}& =& \frac{d}{d r}\bigg( \dot{t}^{-1} \frac{\delta \mathcal{L}}{\delta r}\bigg),\\
A_{2} &= &-(\dot{t} g_{rr})^{-1}.
\end{eqnarray}
For the case of circular orbit, the eigenvalues of the matrix are called
 {principal} Lyapunov exponent which can be written as
\begin{eqnarray}
\label{Lam}
\lambda^2 = A_1 A_2.
\end{eqnarray}
Then the Lagrange's equation of motion leads to
\begin{eqnarray}
\frac{d}{d \tau} \bigg(\frac{\partial \mathcal{L}}{\partial \dot{r}}\bigg)-\frac{\partial \mathcal{L}}{\partial r} = 0,
\end{eqnarray}
and
\begin{eqnarray}
\frac{d}{d \tau} \bigg(\frac{\partial \mathcal{L}}{\partial \dot{r}}\bigg)= \frac{d}{d \tau}(- g_{rr} \dot{r}) = - \dot{r} \frac{d}{d r} (- g_{rr} \dot{r}).
\end{eqnarray}
Thus, the Lyapunov exponent in terms of square of radial velocity $\dot{r}^{2}$, can be expressed as
\begin{eqnarray}\label{Lag}
\frac{\partial \mathcal{L}}{\partial r} &=& \frac{d}{d r}( - \dot{r} g_{rr} )^{2}  \nonumber\\
&=& - \frac{1}{2 g_{rr}} \frac{d}{d r}(\dot{r}^{2} g_{rr}^{2} ).
\end{eqnarray}
Finally, from~(\ref{Lam}) and (\ref{Lag}), the  {principal} Lyapunov exponent can
be written as
\begin{eqnarray}
\label{Lamp}
\lambda^2 = \frac{1}{2} \frac{1}{g_{rr}}\frac{d}{d r}\bigg[\frac{1}{g_{rr}}\frac{d}{d r} (\dot{r} g_{rr})^{2} \bigg].
\end{eqnarray}
For the case of circular geodesics~\cite{Chandrasekhar83}, we have
\begin{eqnarray}
\label{deri}
\dot{r}^{2} = (\dot{r}^{2})' = 0,
\end{eqnarray}
where $\dot{r}^{2}$ is the square of radial potential or effective radial potential.
From~(\ref{Lamp}), we can obtain  the proper time Lyapunov exponent as
\begin{eqnarray}
\label{lap}
\lambda_{p} = \pm \sqrt{\frac{(\dot{r}^{2})''}{2}},
\end{eqnarray}
and the coordinate  {time} Lyapunov~\cite{Cardoso09} can be derived from~(\ref{Lamp})
as follows
\begin{eqnarray}
\label{lac}
\lambda_{c} = \pm \sqrt{\frac{(\dot{r}^{2})''}{2\dot{t}^{2}}}.
\end{eqnarray}
The above Eqs.~(\ref{lap}) and (\ref{lac}) for $\lambda_p$ and $\lambda_c$  {are
respectively } satisfied for any spherically symmetric BH space-times~\cite{pp1,Pugliese11,Setare11}.
Now, we shall drop the $\pm$ sign and consider only positive Lyapunov exponent.
The circular orbit is stable when $\lambda$ is imaginary,  {the}
circular orbit is unstable when $\lambda$ is real and for $\lambda = 0$,
the circular orbit   {becomes} marginally stable or saddle point.  \\

\noindent Due to Pretorius and Khurana~\cite{Pretorius07}, we can define the
critical exponent as
\begin{eqnarray}
\gamma =\frac{ \Omega}{2\pi \lambda} = \frac{T_{\lambda}}{T_{\Omega}},
\end{eqnarray}
where $T_{\lambda}$ represents the Lyapunov time scale, $T_{\Omega}$ represents the orbital
time scale and $\Omega$ represent the angular velocity, where
$T_{\lambda} = \frac{1}{\lambda}$ and $T_{\Omega} = \frac{2\pi}{\Omega}$.
Now, the critical exponent can be written in terms of second order
derivative of the square of radial velocity $(\dot{r}^{2})$, as
\begin{eqnarray}
\gamma_{p} &=& \frac{1}{2\pi} \sqrt{\frac{2\Omega^{2}}{(\dot{r}^{2})''}},\\
\gamma_{c} &=& \frac{1}{2\pi} \sqrt{\frac{2 {\dot{\phi^{2}}}}{(\dot{r}^{2})''}}.
\end{eqnarray}
 {Here, $\phi$ is an angular coordinate}. For circular
geodesics $ (\dot{r}^{2})'' > 0 $, which implies instability.
 {Now, we shall determine the equatorial circular
geodesics of  Hayward space-time}.

 {\section{Equatorial Circular Geodesics  in Spherically Symmetric metric Hayward Space-time}}

In this paper, the metric \cite{Hayward06} for a static, spherically symmetric space-time can be taken as follows
 {\begin{eqnarray}
\label{met}
ds^{2} &=& - f(r) dt^{2} + \frac{dr^{2}}{f(r)} + r^{2} (d\theta^{2} + \sin^{2}\theta d\phi^{2}),~~~~
\end{eqnarray}
where
\begin{eqnarray}
f(r)=  \bigg(1-\frac{2 m r^{2}}{r^{3} + 2l^{2} m} \bigg).
\end{eqnarray}}

Here, the parameters  $l$ and $m$  are positive constants. This is similar
to the Bardeen BH, which can be reduced to the Schwarzschild solution for
$l = 0$, and become flat space-time for m = 0.\\

 { The function $f(r)$ is plotted in Fig.~1. One can observe
that the geometry is similar to the Schwarzschild BH with a single horizon.
Also there could be a single, double or no horizon which depends on the relation
between the parameters.    	Hayward   \cite{Hayward06}  discussed  the  formation  and evaporation  of  this  new  kind  of  regular  and non singular  BH  where  the static region is Bardeen-like and the dynamic regions are Vaidya-like  and this    inspired scientists to construct a compact star model.   This motivates us  for considering the regular Hayward black holes.    }.\\

\begin{figure}
\begin{center}
{\includegraphics[width=0.55\textwidth]{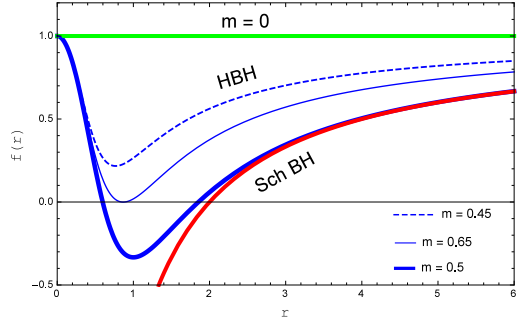}}
\end{center}
\caption{The figure shows the $f(r)$ versus $r$. Here, $l = 0.5$ for Hayward BH.
\label{gm}}
\end{figure}

\subsection{Circular orbits}
To calculate the geodesics in an equatorial plane for space-time of (\ref{met}), we follow
the work of Chandrasekhar \cite{Chandrasekhar83}. In an equatorial plane, we set $\dot{\theta} = 0$ and
$\theta = \frac{\pi}{2}$ = constant.
Here, we restrict our attention to the equatorial orbits, for which the Lagrangian is given by
\begin{eqnarray}
2\mathcal{L} &=& \bigg[- \bigg(1-\frac{2 m r^{2}}{r^{3} + 2l^{2} m} \bigg) \dot{t}^{2}
+ \frac{\dot{r}^{2}}{\bigg(1-\frac{2 m r^{2}}{r^{3} + 2l^{2} m} \bigg)} + r^{2} \dot{\phi}^{2} \bigg],
\end{eqnarray}
where $\phi$ represents an angular coordinate. By using (\ref{Mom}), the generalized
momenta can be represented as
\begin{eqnarray}
\label{pot}
p_{t} &=& - \bigg(1-\frac{2 m r^{2}}{r^{3} + 2l^{2} m} \bigg) \dot{t} = - E = \textrm{const},\\
\label{pof}
p_{\phi} &=& r^{2} \dot{\phi} = L = \textrm{const},\\
\label{pofr}
p_{r} &=& \frac{\dot{r}}{ \bigg(1-\frac{2 m r^{2}}{r^{3} + 2l^{2} m} \bigg)}.
\end{eqnarray}
The Lagrangian is independent on both $t$ and $\phi$, so $p_{t}$ and $p_{\phi}$
are the conserved quantities. Solving Eqs. (\ref{pot}) and (\ref{pof}) for $\dot{t}$ and $\dot{\phi}$, we get
\begin{eqnarray}
\label{dot}
\dot{t} = \frac{E}{\bigg(1-\frac{2 m r^{2}}{r^{3} + 2l^{2} m} \bigg)}~~~~ \textrm{and} ~~~~\dot{\phi} = \frac{L}{r^{2}}.
\end{eqnarray}
The normalization of the four velocity vector $(u^{\alpha})$ can be represented as
an integral equation for the geodesic motion
\begin{eqnarray}
g_{\alpha \beta} u^{\alpha}u^{\beta} = \xi,
\end{eqnarray}
which is equivalent to
\begin{eqnarray}
\label{jai.}
-E \dot{t} + L \dot{\phi} + \frac{\dot{r}^{2}}{\bigg(1-\frac{2 m r^{2}}{r^{3} + 2l^{2} m} \bigg)}  =  \xi.
\end{eqnarray}
Here, $\xi = -1, 0, 1,$ represents the time-like geodesics, null geodesics and space-like geodesics, respectively.
Replacing the values of $\dot{t}$ and $\dot{\phi}$ from (\ref{dot}) in (\ref{jai.}), we obtain the radial
equation  for spherically symmetric space-time:
\begin{eqnarray}
\dot{r}^{2} = E^{2} - \bigg( \frac{L^{2}}{r^{2}} -\xi\bigg)\bigg(1-\frac{2 m r^{2}}{r^{3} + 2l^{2} m} \bigg).
\end{eqnarray}

\subsubsection{Time-like geodesics}
The radial equation of test particle for time-like circular geodesics~\cite{Pradhan56,Pugliese11} is given by
\begin{eqnarray}
\dot{r}^{2} &=& E^{2} - \bigg( 1 + \frac{L^{2}}{r^{2}}\bigg)\bigg(1-\frac{2 m r^{2}}{r^{3} + 2l^{2} m} \bigg).
\end{eqnarray}
For circular orbit with constant $r = r_{\sigma}$ and using the condition (\ref{deri}), we get the energy
and angular momentum per unit mass of the test particle are
\begin{eqnarray}
E_{\sigma}^{2} &=& \frac{(r_{\sigma}^{3} + 2 l^{2} m -2 m r_{\sigma}^{2})^{2}}{(r_{\sigma}^{3}
+ 2 l^{2} m)^{2} - 3 m r_{\sigma}^{5}},\\
L_{\sigma}^{2} &=& \frac{m r_{\sigma}^{4}(r_{\sigma}^{3} - 4 l^{2} m )}{(r_{\sigma}^{3}
+ 2 l^{2} m)^{2} - 3 m r_{\sigma}^{5}}.
\end{eqnarray}
In order to obtain the energy and angular momentum real and finite, the conditions
$(r_{\sigma}^{3} + 2 l^{2} m)^{2} > 3 m r_{\sigma}^{5}$
and $ m r_{\sigma}^{4}(r_{\sigma}^{3} - 4 l^{2} m ) > 0$ must be satisfied.
Thus the orbital velocity becomes
\begin{eqnarray}
\Omega_{\sigma} &=& \frac{\dot{\phi}}{\dot{t}} = \frac{\sqrt{m (r_{\sigma}^{3} - 4 l^{2} m )}}{r_{\sigma}^{3} + 2 l^{2} m}.
\end{eqnarray}

\subsubsection{Null geodesics}
In case of null geodesics, there is no proper time for photons. Thus, we have to calculate only
the coordinate time Lyapunov exponent. The radial equation of the test particle for null circular
geodesics is
\begin{eqnarray}
\label{nullr}
\dot{r}^{2}=  {V_{ef}} = E^{2} - \frac{L^{2}}{r^{2}}\bigg(1-\frac{2 m r^{2}}{r^{3} + 2l^{2} m} \bigg).
\end{eqnarray}
 {The comparison between the Hayward BH and Schwarzschild BH effective potentials
are shown in Fig.~2 for null-circular geodesics. Fig.~2  shows that the effective potential of Hayward
BH is smallest compared to Schwarzschild BH}.

\begin{figure}
\begin{center}
{\includegraphics[width=0.55\textwidth]{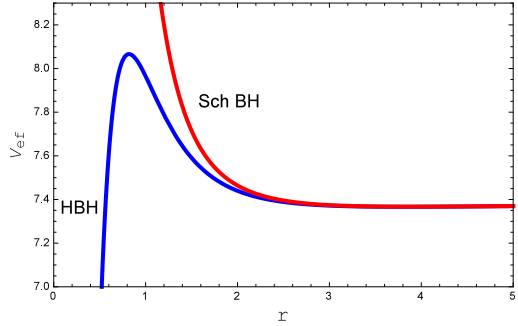}}
\end{center}
\caption{The effective potential $V_{ef}$, for null-circular geodesics in Hayward BH and in
Schwarzschild BH is compared. The constants are with $m = 1, L = 1, E = 1$ and for HBH $l = 0.5$.
\label{gm}}
\end{figure}

Now, the energy and angular momentum at $r = r_{c}$, for the null geodesics is
\begin{eqnarray}
\label{imp}
\frac{E_{c}}{L_{c}} = \pm \sqrt{\frac{(r_{c}^{3} + 2 l^{2} m -2 m r_{c}^{2})}{r_{c}^{2}
(r^{3} + 2 l^{2} m)}}~~~\textrm{and}~~~(r_{c}^{3} + 2 l^{2} m)^{2} - 3 m r_{c}^{5} = 0.
\end{eqnarray}

Let $ D_{c} = \frac{L_{c}}{E_{c}}$ be the impact parameter, then the equation (\ref{imp}) reduces to
\begin{eqnarray}
\frac{1}{ D_{c}} = \frac{E_{c}}{L_{c}} = \frac{\sqrt{m (r_{c}^{3} - 4 l^{2} m )}}{(r_{c}^{3} + 2 l^{2} m)}
= \Omega_{c} =\frac{\dot{\phi}}{\dot{t}}.
\end{eqnarray}

\subsection{Bending of light}
A unstable  {circular} photon orbit is called ``Photon Sphere". The unstable  photon sphere
constitutes  the shadow of the BH.
From  {Eq.~(\ref{pof}) and Eq.~(\ref{pofr}) we find}
\begin{eqnarray}
\label{drphi}
\frac{d r }{d \phi} = \frac{\dot{r}}{\dot{\phi}} = \frac{p_r \bigg(1-\frac{2 m r^{2}}{r^{3} + 2l^{2} m} \bigg) r^2}{L}.
\end{eqnarray}
Again the  {Eq.~(\ref{nullr})} can be written as
\begin{eqnarray}
\label{pr2}
 p_r^2 \bigg(1-\frac{2 m r^{2}}{r^{3} + 2l^{2} m}\bigg) =\frac{ E^2}{(1-\frac{2 m r^{2}}{r^{3} + 2l^{2} m} )} - \frac{L^2}{r^2} .
\end{eqnarray}
We can obtain $p_r$ from above equation as
\begin{eqnarray}
\label{pr}
p_r = \pm  \sqrt{\frac{1}{(1-\frac{2 m r^{2}}{r^{3} + 2l^{2} m} )}} \sqrt{\frac{ E^2}{(1-\frac{2 m r^{2}}{r^{3} + 2l^{2} m} )} - \frac{L^2}{r^2}}.
\end{eqnarray}
Using $p_r$ from  {Eq.~(\ref{pr})}, we can write the  {Eq.~(\ref{drphi})} as follows
\begin{eqnarray}
\label{ndrph}
\frac{d r }{d \phi} = \pm \sqrt{r^2\, \bigg(1-\frac{2 m r^{2}}{r^{3} + 2l^{2} m}\bigg)}
\sqrt{\frac{E^2 }{L^2}\chi^2(r) - 1 },
\end{eqnarray}
where,
 {
\begin{eqnarray}
\chi^2(r) = \frac{r^2}{\left(1-\frac{2 m r^{2}}{r^{3} + 2l^{2} m}\right)}
\end{eqnarray}
}
A light ray which comes in from infinity, reaches at minimum radius $R$
and again goes back to infinity, the bending  {angle} $(\beta_{bending})$ is
given by the formula
 {
\begin{eqnarray}
\beta_{bending} = -\pi +2 \int_R ^\infty \frac{dr}
{\sqrt{r^2 \bigg(1-\frac{2 m r^{2}}{r^{3} + 2l^{2} m}\bigg)\bigg(\frac{E^2 }{L^2}\chi^2(r)-1 \bigg)}}
\end{eqnarray}
Since $R$ is the turning point of the trajectory, the condition $\frac{d r }{d \phi}|_{R = 0}$ must be hold.
Which implies the following equation
\begin{eqnarray}
\chi^2(R) = \frac{L^2 }{E^2}.
\end{eqnarray}
Then the deflection angle can be written in terms of $R$ as
\begin{eqnarray}
\beta_{bending} = -\pi +2 \int_R ^\infty \frac{dr}
{\sqrt{r^2 \bigg(1-\frac{2 m r^{2}}{r^{3} + 2l^{2} m}\bigg)\bigg(\frac{\chi^2(r)}{\chi^2(R)}-1 \bigg)}}
\end{eqnarray}
After putting the value of $\chi^2(r)$ and $\chi^2(R)$, the equation of bending angle of
Hayward BH is
\begin{eqnarray}
\beta_{bending} = -\pi +
2 \int_R ^\infty \frac{dr}
{\sqrt{r^2 \bigg(1-\frac{2 m r^{2}}{r^{3} + 2l^{2} m}\bigg)
\bigg(\frac{r^2}{D^2\left(1-\frac{2 m r^{2}}{r^{3} + 2l^{2} m}\right)}-1 \bigg)}}
\end{eqnarray}
}
where, $D = \frac{L}{E}$ is the impact parameter of the Hayward BH.
 {The exact formula of bending angle is derived in \cite{chiba17} }.

\subsubsection{Radius of the Shadow}
A circular light orbits corresponds to zero velocity and acceleration, so that
$\dot{r} = 0$ and $\ddot{r} = 0$ , implies that $p_r = 0$
and $\dot{p_r} = 0$.
From \ {Eq.~(\ref{pr2})} we obtain
\begin{eqnarray}
\label{pro}
\frac{ E^2}{(1-\frac{2 m r^{2}}{r^{3} + 2l^{2} m} )} - \frac{L^2}{r^2}  = 0
\end{eqnarray}
Now differentiating  {Eq. (\ref{pr2}) with respect to} affine parameter
and  { putting the value of } $p_r = 0$ and $\dot{p_r} = 0$ we have
\begin{eqnarray}
\label{dotpr}
\frac{ E^2(8m^2 l^2 r - 2 m r^4)}{(r^{3} + 2l^{2} m -2 m r^2)^2} + \frac{2 L^2}{r^3}  = 0
\end{eqnarray}
From equations (\ref{pro}) and (\ref{dotpr}) we have
\begin{eqnarray}
L^2 = \frac{ r^2 E^2 (r^{3} + 2l^{2} m)}{(r^{3} + 2l^{2} m-2 m r^2) } = \frac{r^3 E^2 ( 2 m r^4-8m^2 l^2 r )}{2(r^{3} + 2l^{2} m -2 m r^2)^2}
\end{eqnarray}
Subtracting these two equations and after some simplification we can obtain an equation for
radius of the circular light orbit in the following form
 {
\begin{eqnarray}
\label{dotkr}
\frac{d}{dr} \chi^2(r) = 0
\end{eqnarray}
}
Hence, from  {Eq.~(\ref{dotkr}),  the equation
of photon sphere is at $r = r_c$}
\begin{eqnarray}
\label{rps}
 r_{c}^6-3m\,r_{c}^5+4m\,l^2\,r_{c}^3+4m^2\,l^4 &=& 0
\end{eqnarray}
Let $r_{ps}=r_{c}$ be the real root of the equation then $r_{ps}$ is the radius of
circular photon sphere~\cite{chiba17}.
Let $R$ be the critical value of the minimum radius. Then we can write $R$ in terms of $r_{ps}$  as follows
\begin{eqnarray}
R = \frac{r_{ps}^2(r^{3}_{ps} + 2l^{2} m)}{(r^{3}_{ps} + 2l^{2} m-2 m r_{ps}^2)}
\end{eqnarray}
Here, we consider a light ray which is send from the observer's position at $r_{c}$ into the past under an angle $\beta$ with respect to the radial direction. Therefore, we have
\begin{eqnarray}
\cot\beta = \frac{1}{\sqrt{r^2 \bigg(1-\frac{2 m r^{2}}{r^{3} + 2l^{2} m}\bigg)}} \frac{d r }{d \phi}|_{R = 0}
\end{eqnarray}
Again from  {Eq.~(\ref{ndrph})} we have
 {
\begin{eqnarray}
\frac{d r }{d \phi} = \pm \sqrt{r^2 \bigg(1-\frac{2 m r^{2}}{r^{3} + 2l^{2} m}\bigg)} \sqrt{\frac{\chi^2(r)}{\chi^2(R)}- 1}.
\end{eqnarray}
For the angle $\beta$ we have
\begin{eqnarray}
\cot^2\beta = \frac{\chi^2(r_{c})}{\chi^2(R)} - 1,
\end{eqnarray}
and
\begin{eqnarray}
\sin^2\beta = \frac{\chi^2(R)}{\chi^2(r_{c})} = \frac{D^2_c (r^{3}_{c} + 2l^{2} m -2 m r^2_c)}{r_c^2(r^{3}_c + 2l^{2} m)}.
\end{eqnarray}
The boundary of Shadow $\beta_{shadow}$ is described by light rays which spiral asymptotically towards a circular light orbit at radius $r_{ps}$. Then the angular radius of the shadow is given by
\begin{eqnarray}
\sin^2\beta_{shadow} = \frac{\chi^2(r_{ps}) }{\chi^2(r_c)}
= \frac{r^2_{ps} (r^{3}_{ps} + 2l^{2} m -2 m r^2_{ps})(r^{3}_c + 2l^{2} m)}{r_c^2 (r^{3}_{c} + 2l^{2} m -2 m r^2_{c})(r^{3}_{ps} + 2l^{2} m)},
\end{eqnarray}
where $r_{ps}$ has to be determined from the equation (\ref{rps}).
}

\subsection{Lyapunov exponent}
\subsubsection{Time-like case}
Using Eqs. (\ref{lap}) and (\ref{lac}), the proper time Lyapunov exponent and
coordinate time Lyapunov exponent becomes
\begin{eqnarray}
\lambda_{p} &=& \sqrt{\frac{ - m \left[  r_{\sigma}^{5}(r_{\sigma}- 6m)  + 22m\, l^{2}\, r_{\sigma}^{3} -32 l^{4}\, m^{2}\right]}
{(r_{\sigma}^{3} + 2 l^{2}\, m )\left[(r_{\sigma}^{3} + 2m\, l^{2})^{2} - 3 m\, r_{\sigma}^{5}\right]}},\\
\lambda_{c} &=& \sqrt{\frac{-m \left( r_{\sigma}^{5}( r_{\sigma}-6 m) +22m\, l^{2}\, r_{\sigma}^{3} -32\, l^{4}\, m^{2}\right)}
{\left(r_{\sigma}^{3} + 2 m\,l^{2} \right)^{3}}}.
\end{eqnarray}
The time-like circular geodesics is stable when $ r_{\sigma}^{5} (r_{\sigma}-6 m)  +22 l^{2} m r_{\sigma}^{3} -32 l^{4} m^{2}  > 0$, that is, $ \lambda_{p}$ and $\lambda_{c}$ become imaginary. The time-like circular geodesics is
unstable when $  r_{\sigma}^{5} (r_{\sigma}-6 m)  +22 l^{2} m r_{\sigma}^{3} -32 l^{4} m^{2}  < 0$, that is,
$ \lambda_{p}$ and $\lambda_{c}$ become real and the time-like circular geodesics is marginally stable when
$r_{\sigma}^{5}\,(r_{\sigma}-6 m)  +22m\, l^{2}\,r_{\sigma}^{3} -32 l^{4}\, m^{2}  = 0$, that
is, $ \lambda_{p}$ and $\lambda_{c}$  {becomes} zero.

 {Now one can analyze the equation
$r_{\sigma}^6-6 m\,r_{\sigma}^{5}  +22m\, l^{2}\,r_{\sigma}^{3} -32 l^{4}\, m^{2}  = 0$  which
gives us the radii of innermost stable circular orbit. Since it is a non-trivial equation. One can determine
its root numerically for various values of $l$. For example if we choose the value of $l=1$ then one obtains the
ISCO radius at $r_{isco}=5.19m$. For $l=2$, we find ISCO radius is at $r_{isco}=1.91m$.}

\noindent The ratio of proper time Lyapunov exponent and coordinate
time Lyapunov exponent is
\begin{eqnarray}
\frac{\lambda_{p}}{\lambda_{c}} = \frac{(r_{\sigma}^{3} + 2 l^{2} m )}
{\sqrt{(r_{\sigma}^{3} + 2 l^{2} m )^{2}- 3 m r_{\sigma}^{5}}}.
\end{eqnarray}

\noindent One could  see the variation of $\frac{\lambda_{p}}{\lambda_{c}}$
in graphically (See Fig. \ref{gm}) for Hayward BH.

\begin{figure}
\begin{center}
{\includegraphics[width=0.5\textwidth]{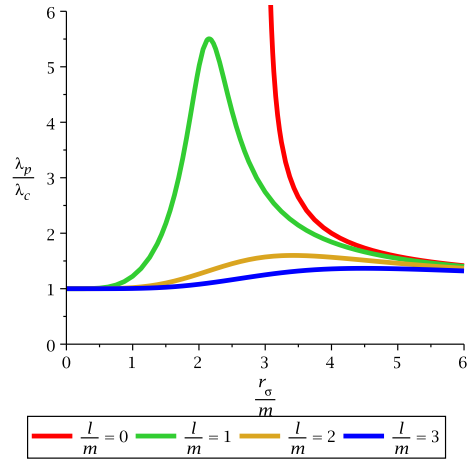}}
\end{center}
\caption{The  variation  of  $\frac{\lambda_{p}}{\lambda_{c}}$ with $\frac{r_{\sigma}}{m}$ for Hayward BH.
\label{gm}}
\end{figure}

 {It can be easily seen from above Fig.\ref{gm} that,  the  ratio of $\lambda_{p}$ and $\lambda_{c}$
varies from orbit to orbit for various values of $\frac{l}{m}$. Also, Fig.\ref{gm}
shows that the ratio $\frac{\lambda_{p}}{\lambda_{c}}$ of Hayward BH is smallest compared
to Schwarzschild ($\frac{l}{m} = 0$) BH}.  \\

\noindent Therefore, the reciprocal of critical exponent is given by
\begin{eqnarray}
\frac{1}{\gamma_{p}} = \frac{T_{\Omega}}{T_{\lambda}}
= 2 \pi \sqrt{\frac{- \left[r_{\sigma}^{5}\,( r_{\sigma}-6 m)
+22m\, l^{2}\, r_{\sigma}^{3} -32 l^{4} m^{2} \right](r_{\sigma}^{3} + 2 l^{2} m )}{(r_{\sigma}^{3} - 4 l^{2} m )
\left[(r_{\sigma}^{3} + 2 l^{2} m )^{2}- 3 m r_{\sigma}^{5}\right]}}.
\end{eqnarray}

\noindent Special case:\\

\noindent For Schwarzschild BH $l = 0 $, the  proper time Lyapunov exponent
and coordinate time Lyapunov exponent are given by
\begin{eqnarray}
\lambda_{p}^{Sch} = \sqrt{\frac{ - m (r_{\sigma}-6 m)}{r_{\sigma}^{3}(r_{\sigma}-3 m)}},\\
\lambda_{c}^{Sch} = \sqrt{\frac{ - m (r_{\sigma}-6 m)}{r_{\sigma}^{4}}}.
\end{eqnarray}
The ratio of $\frac{\lambda_{p}}{\lambda_{c}}$ reduces to
\begin{eqnarray}
\frac{\lambda_{p}}{\lambda_{c}} &=& \sqrt{\frac{r_{\sigma}}{(r_{\sigma}-3 m)}}.
\end{eqnarray}
\noindent The reciprocal of critical exponent for schwarzschild BH is given by
\begin{eqnarray}
\frac{1}{\gamma_{p}} = \frac{T_{\Omega}}{T_{\lambda}} = 2 \pi \sqrt{\frac{ -  (r_{\sigma}-6 m)}{(r_{\sigma}-3 m)}}.
\end{eqnarray}
 {When $r_{\sigma} = 4 m$, the circular orbit become unstable and
critical exponent $(\gamma_{p})$ become $ \frac{1}{2\sqrt{2}\pi}$. In this case,
Lyapunov time scale $T_{\lambda} = \frac{1}{\lambda}$ will be less than the orbital
time scale $T_\Omega = \frac{2\pi}{\Omega}$, that is, $ T_{\lambda}< T_\Omega$ in the
approximation of a test particle around a Schwarzschild BH}.

\subsubsection{Null geodesics}
By using Eq. (\ref{lac}) the Lyapunov exponent for null geodesics is given by
\begin{eqnarray}
\lambda_{c} =  \sqrt{\frac{3m^2 \left[r_{c}^{5} -6 l^{2} ( r_{c}^{3} -2 l^{2} m)\right]}
{(r_{c}^{3} + 2 l^{2} m)^3}}.
\end{eqnarray}

\noindent Here we can see that the circular geodesics is unstable as $ \lambda_{c} $ is real.  \\

\noindent Special case: \\

\noindent For Schwarzschild BH $l = 0 $, the Lyapunov exponent becomes
\begin{eqnarray}
\lambda_{c}^{Sch} = {\frac{\sqrt{3} m}{r_{c}^{2}}}.
\end{eqnarray}
It can be easily seen that for $ r_{c} = 3 m $; $ \lambda_{c}^{Sch}$ is real which
implies that Schwarzschild photon sphere is unstable.

 {\section{Null Circular Geodesic and QNMs for Hayward BH in the Eikonal limit}}
We consider the usual wave like equation, with an effective potential  {which was
first derived } by Iyer and Will~\cite{Iyer87},
\begin{eqnarray}
\frac{d^{2} Y}{d r_{*}^{2}} + \Psi_{0} Y = 0,
\end{eqnarray}
where,
 {\begin{eqnarray}
\Psi_{0} &=& w^2 - V_s(r)\\
\textrm{and}\nonumber\\
V_s(r) &=& \bigg[ \frac{j(j+1)}{r^2} +
\frac{2 m (r^3 -4 l^2 m)}{(r^{3} + 2 l^{2} m)^2} \bigg] \bigg(1 - \frac{2 m r^{2}}{r^{3} + 2 l^{2} m} \bigg)
\end{eqnarray}}
Here,where $ j $ being the angular harmonic index, $ Y $ represents the radial
part of the perturbation variable and $ r_{*} $ is a convenient ``tortoise" coordinate,
ranging from $ -\infty $ to $ +\infty $.  \\

\noindent The radial coordinate $r$ and the tortoise coordinate $ r_{*} $ are
related by the following equation
\begin{eqnarray}
\frac{d r}{ d r_{*}} &=& 1 - \frac{2 m r^{2}}{r^{3} + 2 l^{2} m }.
\end{eqnarray}
In case of the eikonal limit $( j\rightarrow\infty)$, we get
\begin{eqnarray}
\label{psi}
\Psi_{0} \simeq \omega^{2} - \frac{ {j^2}}{r^{2}} \bigg(1 - \frac{2 m r^{2}}{r^{3} + 2 l^{2} m } \bigg).
\end{eqnarray}

\noindent With the help of Eq. (\ref{psi}), we can find the maximum value
of $ \Psi_{0}$ which occurs at $r = r_{o} $
\begin{eqnarray}
(r_{o}^{3} + 2 l^{2} m)^{2} - 3 m r_{o}^{5} = 0.
\end{eqnarray}
Also from the null circular geodesic at $ r = r_{c}$~, we obtain
\begin{eqnarray}
(r_{c}^{3} + 2 l^{2} m)^{2} - 3 m r_{c}^{5} = 0.
\end{eqnarray}
Since the location of the null circular geodesics and the maximum value of
$ \Psi_{0}$ are coincident at $ r_{c} = r_{o} $, then we get the following QNM
condition \cite{Iyer87b,Berti07}
\begin{eqnarray}
\label{in}
\frac{\Psi_{0} (r_{o})}{\sqrt{ - 2 \Psi_{0}^{''}(r_{o})}} = i ( n + 1/2)  ,
\end{eqnarray}
where  $ \Psi_{0}^{''} \equiv \frac{d^{2}\Psi_{0} }{d r_{*}^{2}}$ and Eq. (\ref{in})
is evaluated at an extremum of $  \Psi_{0} $ (the point $r_{0}$ at which $ \frac{d \Psi_{0}}{d r_{*}} = 0$).   \\

\noindent Now, the formula (\ref{in}) allows us to conclude that, in case of the large-$j$ limit
\begin{eqnarray}
\label{qnm}
\omega_{QNM} = j\bigg( \frac{\sqrt{m (r_{c}^{3} - 4 l^{2} m )}}{(r_{c}^{3} + 2 l^{2} m)}\bigg)
- i \left(n + \frac{1}{2}\right) \sqrt{\frac{3 m^2 \left[r_{c}^{5} -6 l^{2} (r_{c}^{3} -2l^{2}\, m)\right]}
{(r_{c}^{3} + 2 l^{2} m )^3}}.
\end{eqnarray}
\begin{figure*}[thbp]
\begin{tabular}{rl}
\includegraphics[width=0.5\textwidth]{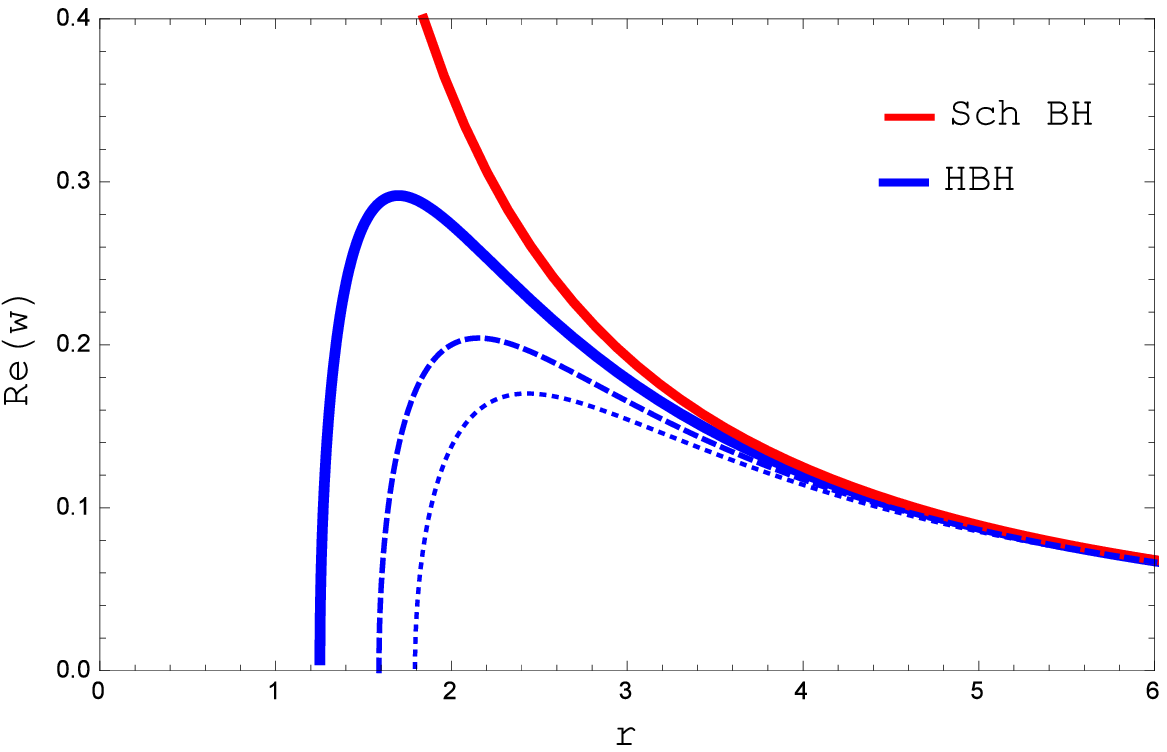}
\includegraphics[width=0.5\textwidth]{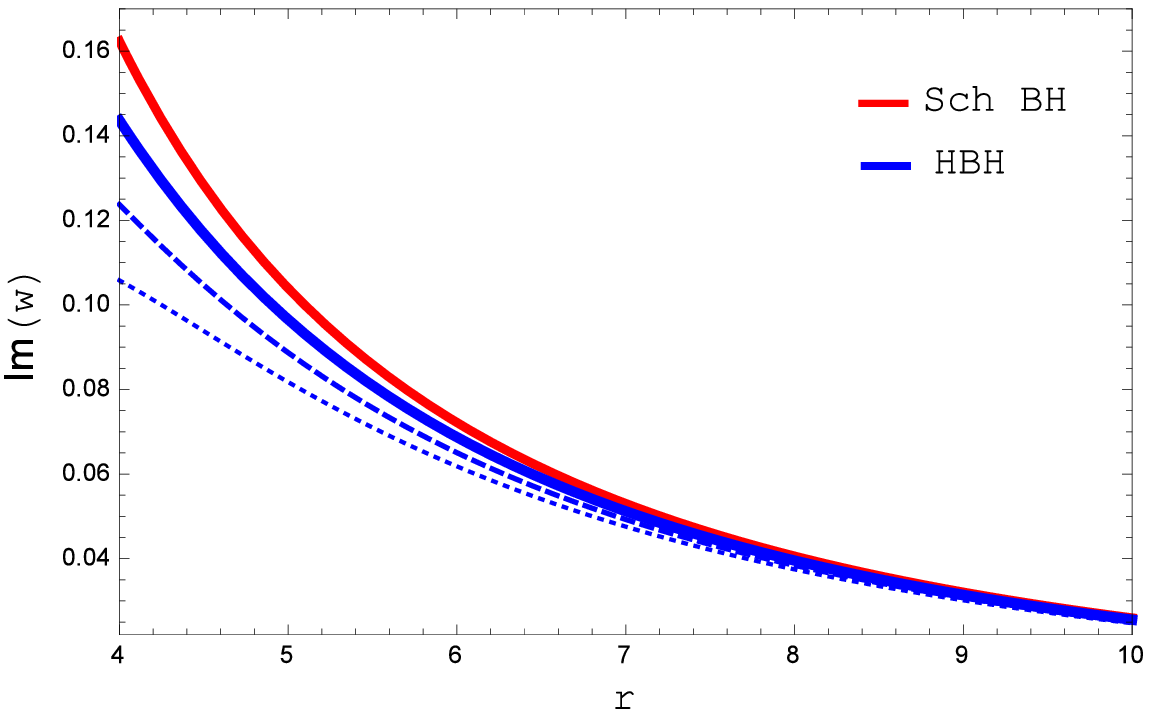}\\
\end{tabular}
\caption{The figure shows QNMs frequency $Re(w)$ versus $ r$ (left Panel) and
$ Im(w)$ versus $r$ (right panel) of HBH and Sch BH ; the other parameters fixed
to $m = 1$ and $ l = 0.7 (\textrm{solid}),l =1 (\textrm{dashed})$ and $l = 1.2 (\textrm{dotted})$ for HBH.}
\label{fvr}
\end{figure*}

\begin{figure*}[thbp]
\begin{tabular}{rl}
\includegraphics[width=0.5\textwidth]{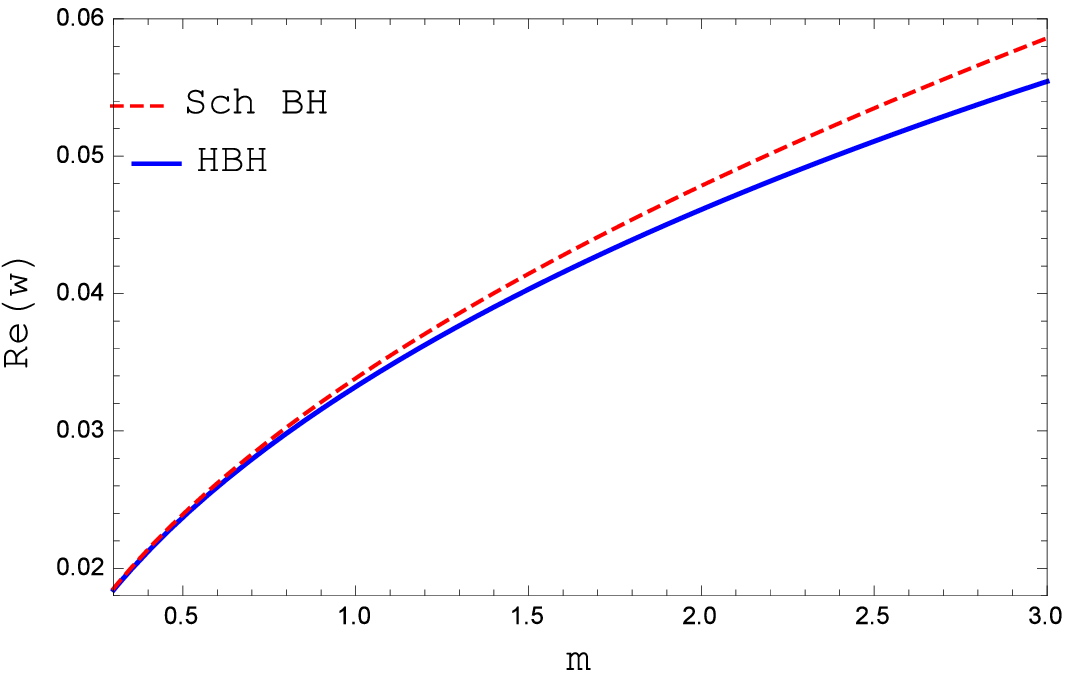}
\includegraphics[width=0.5\textwidth]{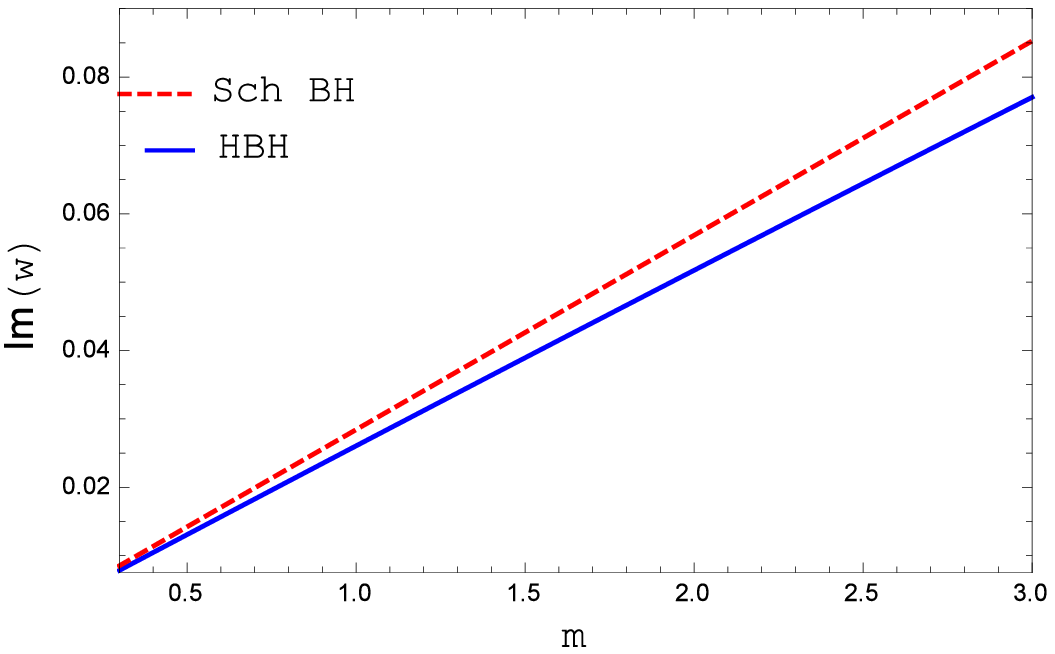}\\
\end{tabular}
\caption{ The figure shows QNMs frequency $Re(w)$ versus $ m$ (left Panel) and
$ Im(w)$ versus $m$ (right panel) of HBH and Sch BH  ; the other parameter fixed to $ l = 0.5$ for HBH. }
\label{fvm}
\end{figure*}

\begin{figure*}[thbp]
\begin{tabular}{rl}
\includegraphics[width=0.5\textwidth]{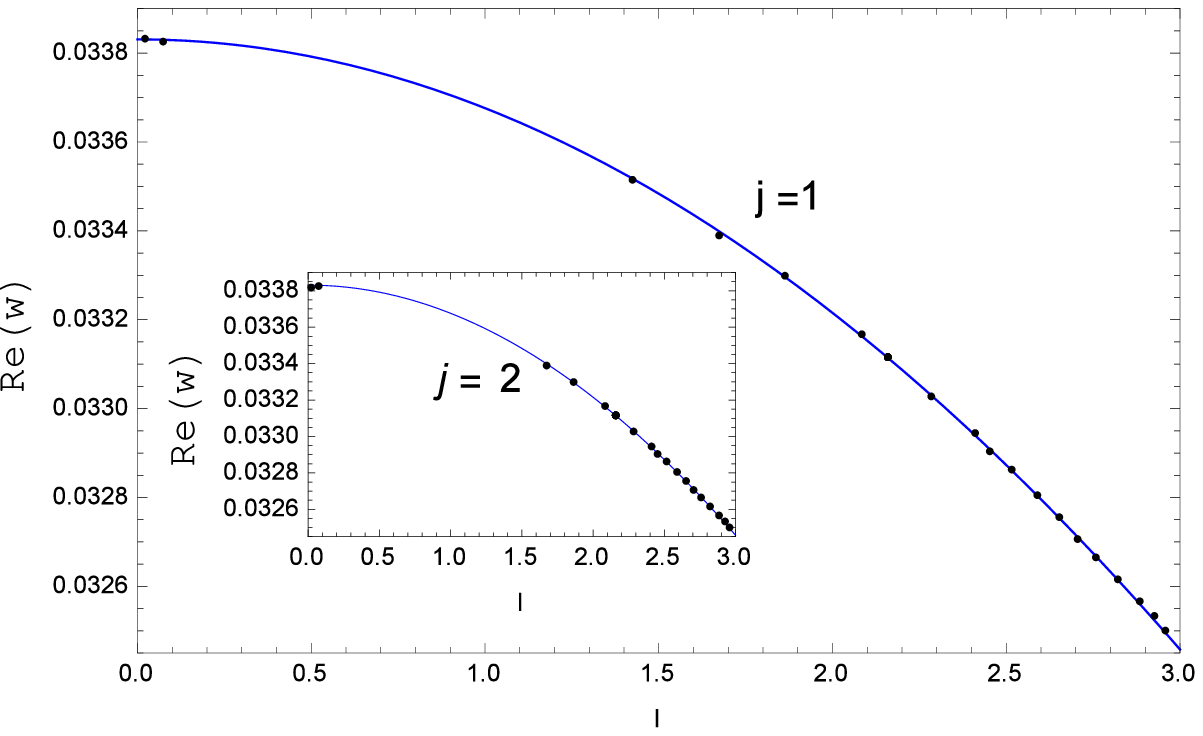}
\includegraphics[width=0.5\textwidth]{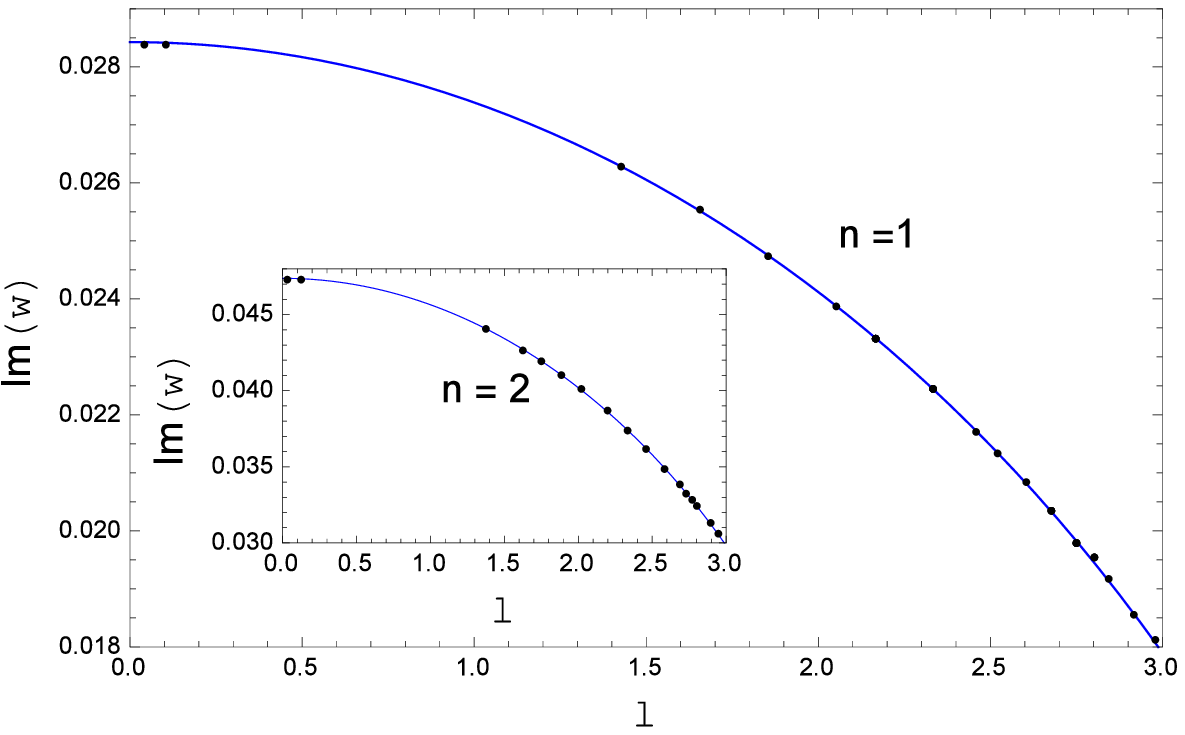}\\
\end{tabular}
\caption{ The figure shows QNMs frequency $Re(w)$ versus $ l$ (left Panel) and
$ Im(w)$ versus $l$ (right panel) of Hayward BH; the other parameter fixed to $m =1$ .}
\label{fvl}
\end{figure*}
 {Cardoso et al.~\cite{Cardoso09} showed in general sense that  this is
one of the most important results of QNMs and the significance of the Eq.~(\ref{qnm})
is that in case of eikonal limit \cite{Baker08}, the real and imaginary parts of the
QNMs \cite{Kokkotas99,Nollert99,Yadav20,Tangherlini63} of the spherically
symmetric, asymptotically flat Hayward regular BH space-time are given by the
frequency and instability time scale of the unstable null circular geodesics}. \\

\noindent Special case:  \\

\noindent For Schwarzschild BH l = 0, in case of eikonal limit the frequency of QNM is given by
\begin{eqnarray}
\omega_{QNM} = j\, \sqrt{\frac{m}{r_{c}^{3}}} - i \left(n +\frac{1}{2} \right)\frac{\sqrt{3}m}{r_{c}^2}.
\end{eqnarray}
Hence, by determining the Lyapunov exponent, we established that in case of eikonal limit, the
frequency of quasi-normal modes of Schwarzschild BH might be determined by the parameters
of the null circular geodesics.\\

In Fig.\ref{fvr},  { Re(w) is plotted as a function of  $r$ (left panel)
by varying $l$. When $l$ increase, the height of the $Re(w)$ decrease. Also $Im(w)$ is
plotted as a function of  $r$ (right panel) by varying $l$.  One can observe that $Im(w)$
decreases when $r$ increases and $Im(w)$ of Hayward BH is the smallest compared to Schwarzschild BH}.\\

 The Fig.\ref{fvm},  {shows that the behavior is similar for Hayward BH and
 Schwarzschild BH in both cases $Re(w)$ and $Im(w)$ increase when $m$ is increase, respectively.
 In fact $Im(w)$ is linearly dependent on $m$.  Also, the angular velocity and instability time
 scale of null-circular geodesics of Hayward BH is the smallest compared to Schwarzschild BH
 regardless of the value of mass, respectively}.\\

In Fig.\ref{fvl},  { $Re(w)$ is plotted against  $l$ (left panel) and $Im(w)$ is
plotted against $l$ (right panel). Both  of $Re(w)$ and $Im(w)$ are decrease when $l$ increases,
that is, the modes decays faster  for large $l$. Compared to the Schwarzschild BH ($ l = 0$) the
modes decays faster}.

\section{Conclusions}
 {We investigated the geodesic stability via Lyapunov exponent
for a  static, spherically symmetric regular Hayward BH. We have considered both
time-like case and null case. Using Lyapunov exponent one can easily
determined  whether the geodesics is stable or unstable or marginally stable.
Also, we computed the QNMs frequency in the geometric-optics approximation~(eikonal)
limit. We have showed that the real part of QNMs frequency are evaluated by the
angular velocity in terms of the  unstable circular
photon orbit. While the imaginary part is related to the instability time scale of
photon orbit.}

 {Moreover, we  examined  the Lyapunov exponent that could be used
to established the instability of equatorial circular geodesics both for time-like and null
cases. When the parameter $l=0$, one gets the result of Schwarzschild  BH.
We derived both  the proper time Lyapunov exponent and coordinate time
Lyapunov exponent.  We have calculated their ratio.  We have also calculated the
reciprocal of  critical exponent. Moreover, we have
showed that for any unstable circular orbit, $ T_{\Omega} > T_{\lambda}$, that is,
orbital time scale is greater than the Lyapunov time scale.}

 {The most important result that we derived is the relation
between unstable null circular geodesics and QNMs frequency in the eikonal
limit. Moreover we discussed the  gravitational bending of light and radius of shadow
for this regular BH.}

\section*{Conflict of interest}
The authors declare that they have no conflict of interest.

\section*{Acknowledgements}

FR  would like to thank the authorities of the Inter-University Centre for Astronomy and Astrophysics, Pune,
India for providing the research facilities.  FR is  also thankful to DST-SERB, Govt. of India and RUSA 2.0,
Jadavpur University  for financial support. Finally we are grateful to the referees for their valuable comments and suggestions.


\end{document}